\documentclass[10pt,preprintnumbers,showpacs,amsmath,amssymb,cite]{revtex4}
\input epsf.sty
\usepackage{txfonts}
\usepackage{bm}   
\usepackage{graphicx,amsmath,mathrsfs}
\usepackage{multirow}

\newcommand{\nn}{\nonumber}  

\begin{document}

\title{\bf Extend the Kompaneets Equation to Down-Comptonization Process in a Novel Way}

\author{
{\bf Xu Zhang}$^{a,b}$, {\bf Xurong Chen}$^a$
\\
\normalsize $^a$Institute of Modern Physics, Chinese Academy of Sciences, Lanzhou 730000, P.R. China\\
\normalsize $^b$University of Chinese Academy of Sciences, Beijing 100049, P.R. China \\
}

\date{\today}

\begin{abstract}

Comptonization is a very important phenomenon in astrophysics. Kompaneets equation describes
up-Comptonization process in nonrelativistic astrophysics, while it fails for down-Comptonization Scattering which is the most
important radiative transfer process in hard X-rays and $\gamma$-rays astronomy. In this study,
we explore both up-Comptonization and down-Comptonization processes. A new
relativistic corrections is introduced to the Kompaneets equation which is valid in both nonrelativistic
energy regime with the photon energy $h\nu << m_ec^2$ and the electron temperature $KT_e << m_ec^2$.
Numerical comparisons are presented which show excellent agreement between the Kompaneets equation
and the new equation.


\end{abstract}



\footnotetext[0]{{\bf Corresponding author:} \rm Xurong Chen, E-mail: xchen\textcircled{a}impcas.ac.cn. }

\maketitle

\section{Introduction}   

The radiative transfer for photons that results from electron scattering-plays a crucial role
in determining the spectra that emerge from cosmic X-ray and $\gamma$-ray sources, and that
has been an important topic in both astrophysics and radiation physics. With the rapid
development of X-ray astronomy in recent years, it seems necessary to improve the available
theory for the study of the radiative transfer process.

Comptonization is of vital importance in astro-particle physics. When the
high energy photons, usually X-rays or Gamma-rays, from the star or some other celestial
bodies are going through the media surrounding them, these media would be ionized into
plasma. Comptonization is just the interaction process between the high energy photons
and the plasma.

In Comptonization, there would exist many Thomson scatterings, many Compton scatterings
and many electron-position pair creations as well. Since the Thomson scattering can’t
change the photons energy, the Compton scattering is the majoring process in non-relativistic
Comptonization. A mono-Compton scattering is just a collision between photo and electron,
so we have kinetic energy and momentum energy conserved.

The various studies of the radiative
transfer for photons have employed Monte Carlo calculations$^{[1-3]}$ and solution of the
Kompaneets equation$^{[4]}$. Monte Carlo calculations is cumbersome and poorly suited for
studying the nonlinear problem$^{[5-6]}$, e.g. the Bose-Einstein spectrum. With
the Kompaneets equation one can handle nonlinear problem.

The Kompaneets equation is particular form of a Fokker-Plank equation.
This equation as follows,
\begin{align}
\label{eq1}
(\frac{\partial{n}}{\partial{t}})_c=\frac{KT_e}{m_ec^2}N_e\sigma_Tc\frac{1}{x^2}\frac{\partial}{\partial{x}}\{x^4[\frac{\partial{n}}{\partial{x}}+n(n+1)]\}
\end{align}
where x$\equiv h\nu/$kTe is the dimensionless photon energy; N$_e$ is the number density
of the scattering electron gas; $\sigma_T$ is the Thomson cross-section; n(x,t) $\equiv$ n($\nu$,t)
is the frequency distribution function of the photon gas.
The Kompaneets equation describes the up-Comptonization Scattering of low energy photons of frequency $\nu$ on a dilute
distribution of nonrelativistic electrons when all photons and electrons are distributed isotropically
in their momenta. And the Kompaneets equation is applied with the photon energy $h\nu << m_ec^2$ and
the electron temperature $h\nu << KT_e$. While it fails for use in down-Comptonization of
high energy photons passing through electron plasma which is the most important radiative
transfer process in hard X-rays and $\gamma$-rays astronomy.

Based on Fokker-Plank equation, Ross and McCray~\cite{ross1978} obtained the Ross-McCray equation
to describe Compton Scattering:
\begin{align}
\label{eq2}
(\frac{\partial{n}}{\partial{t}})_c=&\frac{KT_e}{m_ec^2}N_e\sigma_Tc \frac{1}{x^2}\frac{\partial}{\partial{x}}\{x^4[n+(1+\frac{7}{10}\frac{KT_e}{m_ec^2}x^2)\frac{\partial{n}}{\partial{x}}]\}
\end{align}
For correct diffusion equation, when photon-gas reaches a thermal equilibrium with
the electrons, $\frac{\partial{n}}{\partial{t}}=0$ should be satisfied.
While inserting Plank distribution function $n(x)=(e^x-1)^{-1}$ to Eq.(2), $\frac{\partial{n}}{\partial{t}}\neq0$.

Liu et al. extended the Kompaneets equation~\cite{liu2004} which aims at describing a more general Compton scattering process,
\begin{align*}
(\frac{\partial{n}}{\partial{t}})_c=&\frac{KT_e}{m_ec^2}N_e\sigma_Tc\frac{1}{x^2}\frac{\partial}{\partial{x}}\{x^4(1+\frac{7}{10}\frac{KT_e}{m_ec^2}x^2)[\frac{\partial{n}}{\partial{x}}+n(n+1)]\}\tag{3}
\end{align*}

Following Kompaneets, they assumed that $\Delta{\nu}$
is also the small quantity when $h\nu >> KT_e$, and used $\Delta{\nu}$ expanding the distribution
function. While when $h\nu >> KT_e$, assuming $\Delta{\nu}$ as the small quantity is
not strictly satisfied. The change of the photon energy in each collision
is given by the following well known formula (if the electron is approximately
motionless compared with photon before collision, $h\nu >> KT_e$)
\begin{align*}
h\Delta{\nu}=hv\frac{\lambda (1-cos\theta)}{1+\lambda(1-cos\theta)}\qquad  \lambda=\frac{hv}{m_0c^2}\tag{4}
\end{align*}
The change of the photon energy depends on $h\nu$ and the scattering angle. With the
increasing of the scattering angle, $h\Delta \nu$ also increases. When scattering
angle reaches to $\pi$ , $h\Delta \nu$ reaches to its maximum value. \\
For example:
\begin{itemize}
  \item h$\nu$ =1 KeV, h$\Delta{\nu}_{max}$ = 0.0039 KeV;
  \item h$\nu$ =10 KeV, h$\Delta{\nu}_{max}$ = 0.38 KeV;
  \item h$\nu$ =30 KeV, h$\Delta{\nu}_{max}$ = 3 KeV.
\end{itemize}

So in the down-Comptonization process $KT_e << h\nu$, the condition $h\Delta{\nu} << KT_e$ is not always satisfied.

The purpose of this paper is to re-derive the Kompaneets equation in a novel way. We introduce
relativistic corrections to Kompaneets equation to describe radiative transfer process
in hard X-rays and $\gamma$-rays astronomy. While different from Kompaneets, we chose
the change of the electron's momentum $|\Delta \vec{p}|$ expanding the distribution function.
In the following this paper, we can see $\Delta{p}$ is more approximately to the small quantity than $\Delta{\nu}$  both
when $h\nu >> KT_e$ and when $h\nu << KT_e$.

In Section 2, we will obtain a new relativistic corrections to Kompaneets equation in a novel way.
Then in Section 3 we compare our new equation to the Kompaneets equation and Liu's equation.

\section{Extentions to the Kompaneets equation}

Following Kompaneets, we treat the radiation as a closed system that consists of photon-gas and electron-gas.
Although the system can't be described by a characteristic temperature before the thermal equilibrium being
established, the electron-gas itself is already in thermal equilibrium as the interaction between electrons
is the Coulomb long-range force. For a tenuous radiations field, Fermi distribution of electron-gas
approximate as Boltzman distribution
\begin{align*}
f(p)=f_0exp[-\frac{1}{KT_e}\frac{p^2}{2m_e}]\tag{5}
\end{align*}
Considering photon which is boson that is in the system, which doesn't appreciably disturb
the field that electron-gas has established, but which is capable of absorption and radiation
of energy of all frequencies. Over a sufficiently long time, the absorption and radiation of
the photon-gas caused by the scattering of electron and photon will lead to thermal equilibrium.

In the electron and photon collisions process that $( \vec{p},\vec{v},\vec{n})\rightarrow ( \vec{p'},\vec{v'},\vec{n'})$
leads to a decrease of the photon number$n(\nu,t)$, as describing the number of electrons
in the interval $\vec{p}-\vec{p}+d\vec{p}$ is $N_ef(\vec{p})d^3\vec{p}$, because the photon is boson,
the total transition number in unit volume is
\begin{align*}
n(1+n')N_ef(\vec{p})d^3\vec{p}dw\tag{6}
\end{align*}

The inverse process that $( \vec{p'},\vec{\nu'},\vec{n'})\rightarrow ( \vec{p},\vec{v},\vec{n})$
leads to a increase of the photon number$n(v,t)$,  the total transition number in unit volume is
\begin{align*}
n'(1+n)N_ef(\vec{p'})d^3\vec{p'}dw\tag{7}
\end{align*}
Where $n=n(v,t)$, $n'=n(v',t)$ are the photon numbers before and after collision. And $\vec{p}$, $\vec{p'}$
are the electron momentum before and after collision. The transition probability $dw$ is same as
in collisions process $( \vec{p},\vec{v},\vec{n})\rightarrow ( \vec{p'},\vec{v'},\vec{n'})$ and
inverse process that $( \vec{p'},\vec{v'},\vec{n'})\rightarrow ( \vec{p},\vec{v},\vec{n})$,
because in the non-relativistic limit, the Compton differential scattering cross section
can be approximately expressed by the Thomson section

\begin{align*}
dw=cd\sigma_T=c\frac{r_0^2}{2}(1+cos^2{\theta})2\pi{sin{\theta}}{d}\theta\tag{8}
\end{align*}

the Klein-Nishina cross-section $\frac{d\sigma_T}{{d}\theta}$ has the same value both
for the scattering angle ${\theta}$ and $\pi-{\theta}$.
Therefore the change of the distribution function $n(\nu,t)$ that result from photon
and electron scattering is

\begin{align*}
(\frac{\partial{n}}{\partial{t}})_c=&-N_e \int d^3\vec{p'}\int [n(1+n')f(\vec {p'})-n'(1+n)f(\vec {p'})]dw \tag{9}
\end{align*}

The law of energy and momentum conservation in the non-relativistic approximation are written in the form
\begin{align*}
\frac{hv}{c}\vec{n}+\vec{p}&=\frac{hv}{c}\vec{n'}+\vec{p'}\nn\\
                   hv+\frac{p^2}{2m_e}&=hv'+\frac{p'^2}{2m_e}\tag{10}
\end{align*}
Where $\vec{n}$ and $\vec{n'}$ are the direction of photon before and after collision, respectively.
The expression $\Delta{v}=v-v', \Delta \vec{p}=\vec{p}-\vec{p'}$ can be obtained from Eq.(10).
Retaining only the first order of $\Delta{v}$, we obtain
\begin{align*}
h\Delta v=&-[\frac{hvc\vec{p}}{m_ec^2}(\vec{n}-\vec{n'})+\frac{(hv)^2}{m_ec^2}(1-\vec{n}\vec{n'})]\tag{11}\nn\\
           h\Delta v=&-\frac{1}{2m_e}[2|\vec{p}||\Delta \vec{p}|cos\gamma+|\Delta \vec{p}|^2]\tag{12}\nn\\
           {|\Delta\vec{p}|^2}=&(\frac{hv}{c})^2(\vec{n}-\vec{n'})^2-2(\vec{n}-\vec{n'})\vec{n'}(\frac{hv}{c})\frac{h\Delta{v}}{c}\nn\\
           &+(\frac{h\Delta v}{c})^2\tag{13}
\end{align*}
Where $\gamma$ is the angle between $\vec{p}$ and $\Delta\vec{p}$. For convenience,
in the following calculation, we use $\Delta{p}$ replacing $|\Delta \vec{p}|$.

By expanding $n'=n(v',t)$ and $f(\vec{p'})$ in $\Delta{p}$ to second order,
and replacing the frequency $\nu$ by a convenient dimensionless frequency $x$, where $x=\frac{hv}{KT_e}$, we obtain
\begin{align*}
 n'=&n+\frac{\partial{n}}{\partial{x}}[-\frac{1}{{KT_e}}\frac{1}{2m_e}{2{p}cos\gamma}\Delta {p}+
(-\frac{1}{KT_e})\frac{1}{2m_e}(\Delta {p})^2]+\frac{1}{2}\{\frac{\partial^2{n}}{\partial{x}^2}[-\frac{1}{{KT_e}}\frac{1}{2m_e}{2{p}cos\gamma}\Delta {p}]^2\tag{14}
\end{align*}

\begin{align*}
    f(p')=&f_0Exp[-\frac{1}{KT_e}\frac{1}{2m_e}(\vec{p}+\Delta \vec{p})^2]\nn\\
     =&f(p)+f(p)[-\frac{1}{{KT_e}}\frac{1}{2m_e}{2{p}cos\gamma}\Delta {p}-\frac{1}{KT_e}\frac{1}{2m_e}(\Delta {p})^2]+\frac{1}{2}f(p)[-\frac{1}{{KT_e}}\frac{1}{2m_e}{2{p}cos\gamma}\Delta{p}]^2\tag{15}
\end{align*}

The first order is same as expanding $n'=n(\nu',t)$ and $f(\vec{p'})$ in terms of $\Delta{\nu}$.
While there are difference for the second and higher orders.

The Compton Scattering and the inverse Compton Scattering consist of the whole scattering process of photon and electron.
In Compton Scattering, which is satisfied $h\Delta{\nu}<0$. Using Eq.(12) and (13), we obtain
\begin{align*}
-\frac{1}{2m_e}2|\vec{p}||\Delta \vec{p}|cos\gamma=&
-[\frac{hvc\vec{p}}{m_ec^2}(\vec{n}-\vec{n'})]+(1-\vec{n}\vec{n'})(\frac{hv}{m_ec^2}){h\Delta v}+\frac{(h\Delta v)^2}{2m_ec^2}\tag{16}
\end{align*}
The last term is a small quantity compared to other terms, therefore can be ignored.

As the second term
$(1-\vec{n}\vec{n'})(\frac{h\nu}{m_ec^2}){h\Delta \nu}<0$, we obtain
\begin{align*}
|-\frac{1}{2m_e}2|\vec{p}||\Delta \vec{p}|cos\gamma|<|-[\frac{hvc\vec{p}}{m_ec^2}(\vec{n}-\vec{n'})]|\tag{17}
\end{align*}

If the electron is approximately motionless compared with photon before collision $(h\nu >> KT_e)$,
we obtain
\begin{align*}
|-[\frac{hvc\vec{p}}{m_ec^2}(\vec{n}-\vec{n'})]|<<\frac{(hv)^2}{m_ec^2}(1-\vec{n}\vec{n'})]\tag{18}\nn\\
|-\frac{1}{2m_e}2|\vec{p}||\Delta \vec{p}|cos\gamma|<<h\Delta v\nn\tag{19}
\end{align*}
If the thermal energy of electrons is markedly larger than the energy of photons in radiation field$(KT_e>>hv)$:
\begin{align*}
|-[\frac{hvc\vec{p}}{m_ec^2}(\vec{n}-\vec{n'})]\approx h\Delta v\nn\\
|-\frac{1}{2m_e}2|\vec{p}||\Delta \vec{p}|cos\gamma|<h\Delta v\nn\tag{20}
\end{align*}

So we have proved in Compton Scattering, $\Delta{p}$ is a smaller quantity
than $\Delta{\nu}$ in both cases $h\nu >> KT_e$ and $h\nu << KT_e$.

In inverse Compton Scattering which only happened when low energy photon collides
with high energy electron$(KT_e > h\nu)$ that is satisfied $h\Delta{\nu} > 0$.

from Eq.(11), we obtain $|c\vec{p}(\vec{n}-\vec{n'})|>|(1-\vec{n}\vec{n'})hv|$. So that $|c\vec{p}(\vec{n}-\vec{n'})|>>|(1-\vec{n}\vec{n'})h\Delta{v}|$.
Therefore Eq.(16) can be be approximately to
\begin{align*}
-\frac{1}{2m_e}2|\vec{p}||\Delta \vec{p}|cos\gamma\approx-[\frac{hvc\vec{p}}{m_ec^2}(\vec{n}-\vec{n'})]\approx h\Delta{v}  \tag{21}
\end{align*}

Eq.(21) is what Kompaneets used when he derived his equations in up-Comptonization
process which is applied to non-relativistic astrophysics problems. Here we have proved $\Delta{p}$
is a smaller quantity than $\Delta{\nu}$ in both cases $h\nu >> KT_e$ and $h\nu << KT_e$.

So we expand $n'=n(v',t)$ and $f(\vec{p'})$ in terms of $\Delta{p}$ to higher order:
\begin{align*}
 n'=&n+\frac{\partial{n}}{\partial{x}}[-\frac{1}{{KT_e}}\frac{1}{2m_e}{2{p}cos\gamma}\Delta {p}+
(-\frac{1}{KT_e})\frac{1}{2m_e}(\Delta {p})^2]+\frac{1}{2}\frac{\partial^2{n}}{\partial{x}^2}\{[-\frac{1}{{KT_e}}\frac{1}{2m_e}{2{p}cos\gamma}\Delta {p}]^2\nn\\
  &+\frac{1}{3}(\frac{1}{KT_e})^2(\frac{1}{m_e})^2{p}cos\gamma(\Delta {p})^2\}+\frac{1}{6}\frac{\partial^3{n}}{\partial{x}^3}[-\frac{1}{{KT_e}}\frac{1}{2m_e}{2{p}cos\gamma}\Delta {p}]^3\tag{22}\nn\\
    f(p')=&f_0Exp[-\frac{1}{KT_e}\frac{1}{2m_e}(\vec{p}+\Delta \vec{p})^2]\nn\\
     =&f(p)+f(p)[-\frac{1}{{KT_e}}\frac{1}{2m_e}{2{p}cos\gamma}\Delta {p}-\frac{1}{KT_e}\frac{1}{2m_e}(\Delta {p})^2]\nn\\
      &+\frac{1}{2}f(p)\{[-\frac{1}{{KT_e}}\frac{1}{2m_e}{2{p}cos\gamma}\Delta{p}]^2+\frac{1}{3}(\frac{1}{KT_e})^2(\frac{1}{m_e})^2{p}cos\gamma(\Delta {p})^2\}\nn\\
      &+\frac{1}{6}f(p)[-\frac{1}{{KT_e}}\frac{1}{2m_e}{2{p}cos\gamma}\Delta {p}]^3\tag{23}
\end{align*}

Insert Eq.(14) and Eq.(15) into the Eq.(9), we obtain
\begin{align*}
     (\frac{\partial{n}}{\partial{t}})_c& = N_e[\frac{\partial{n}}{\partial{x}}+n(n+1)]\int{d^3\vec{p}}\int{dW}f(p)\{[-\frac{1}{{KT_e}}\frac{1}{2m_e}{2{p}cos\gamma}]\Delta {p}+\nn\\
     &(-\frac{1}{KT_e})\frac{1}{2m_e}(\Delta {p})^2\}+\frac{N_e}{2}[\frac{\partial^2n}{\partial{x^2}}+2(n+1)\frac{\partial{n}}{\partial{x}}+n(n+1)]\times \nn\\
                                        &\int{d^3\vec{p}}\int{dW}f(p))\{[-\frac{1}{{KT_e}}\frac{1}{2m_e}{2{p}cos\gamma}]^2(\Delta {p})^2\}\tag{24}
\end{align*}

Next, Follow Kompaneets' method, we first calculate the second integral of the Eq.(24).
The other is determined from the condition that the equation ought to guarantee conservation of
the total number of quanta in the scattering.Let:
\begin{align*}
     I=\int{d^3p}\int{dW}f(p))[h\Delta \nu +\frac{1}{2m_e}|\Delta \vec{p}|^2]^2\tag{25}
\end{align*}
Inserting Eq.(13) into the Eq.(25), we obtain
\begin{align*}
     I=&\int{d^3\vec{p}}\int{dW}f(p))\{(h\Delta \nu)^2+\frac{1}{4m_e^2}[(\frac{hv}{c})^2(\vec{n}-\vec{n'})^2\nn\\
     &-2(\vec{n}-\vec{n'})\vec{n'}(\frac{h\nu}{c})\frac{h\Delta \nu}{c}+(\frac{h\Delta \nu}{c})^2]^2+\frac{1}{m_e}(h\Delta \nu)[(\frac{h\nu}{c})^2(\vec{n}-\vec{n'})^2-2(\vec{n}-\vec{n'})\vec{n'}(\frac{h\nu}{c})\frac{h\Delta v}{c}+(\frac{h\Delta \nu}{c})^2]\}\tag{26}
\end{align*}

Let:
\begin{align*}
I_{1}=&\int{d^3\vec{p}}\int{dW}f(p))(h\Delta \nu)^2 \nn\\ I_{2}=&\int{d^3\vec{p}}\int{dW}f(p)\frac{1}{4m_e^2}[(\frac{h\nu}{c})^2(\vec{n}-\vec{n'})^2\nn\\
&-2(\vec{n}-\vec{n'})\vec{n'}(\frac{h\nu}{c})\frac{h\Delta \nu}{c}+(\frac{h\Delta \nu}{c})^2]^2\nn\\
I_{3}=&\int{d^3\vec{p}}\int{dW}f(p)\frac{1}{m_e}(h\Delta \nu)[(\frac{h\nu}{c})^2(\vec{n}-\vec{n'})^2-2(\vec{n}-\vec{n'})\vec{n'}(\frac{h\nu}{c})\frac{h\Delta \nu}{c}+(\frac{h\Delta \nu}{c})^2]
\end{align*}

Calculating those integral above, we obtain
\begin{align*}
     I_{1}&=[\frac{2KT_e}{m_ec^2}(h\nu)^2\sigma_Tc+\frac{(h\nu)^4}{m_e^2c^4}\frac{7}{5}\sigma_Tc]\nn\\ I_{2}&=\frac{(h\nu)^4}{m_e^2c^4}\frac{7}{5}\sigma_Tc\nn\\
     I_{3}&=-\frac{(h\nu)^4}{m_e^2c^4}\frac{14}{5}\sigma_Tc+\frac{28}{5}\frac{KT_e}{(m_ec^2)^2}(h\nu)^3\sigma_Tc
\end{align*}

Hence we can get
\begin{align*}
I=\frac{2KT_e}{m_ec^2}(h\nu)^2\sigma_Tc+\frac{28}{5}\frac{KT_e}{(m_ec^2)^2}(h\nu)^3\sigma_Tc\nn
\end{align*}

The Eq.(24) obeys a sort of conservation laws:
\begin{align*}
(\frac{\partial{n}}{\partial{t}})_c=-x^{-2}\frac{\partial(x^2j)}{\partial{x}}\tag{27}\nn
\end{align*}
Where $j$ is the flow of quanta in the frequency space. Using spherical coordinates $({\chi},{\theta},{\varphi} )$ to replace $(x_1, x_2, x_3)$, Eq.(27) can be written as
\begin{align*}
(\frac{\partial{n}}{\partial{t}})_c=-\frac{2}{x}j-\frac{\partial{j}}{\partial{x}}\tag{28}\nn
\end{align*}

The Eq.(24) is of second order relative to $x$, and dependent on the second derivative $\frac{\partial^2{n}}{\partial{x^2}}$
linearly, so that the current must contain the first derivative $\frac{\partial{n}}{\partial{x}}$.
Meanwhile in the state of thermal equilibrium, the distribution function is Planckian, $n(x)=(e^x-1)^{-1}$,
the flow vanishes.

Thus $\frac{\partial{n}}{\partial{x}}=-n(n+1)$.

Therefore $j$ follows the form:
\begin{align*}
j(x)=g(x)[\frac{\partial{n}}{\partial{x}}+n(n+1)]\nn\tag{29}
\end{align*}
Inserting Eq.(29) into the Eq.(28), we obtain
\begin{align*}
(\frac{\partial{n}}{\partial{t}})_c=&-g(x)[\frac{\partial^2n}{\partial{x^2}}+2(n+1)\frac{\partial{n}}{\partial{x}}]+[\frac{\partial{g}}{\partial{x}}+\frac{2g}{x}][\frac{\partial{n}}{\partial{x}}+n(n+1)]\tag{30}
\end{align*}
Meanwhile the Eq.(24) can be written as
\begin{align*}
(\frac{\partial{n}}{\partial{t}})_c=&N_e\frac{1}{KT_e}[\frac{\partial{n}}{\partial{x}}+n(n+1)]H + \frac{N_e}{2}(\frac{1}{KT_e})^2[\frac{\partial^2n}{\partial{x^2}}+2(n+1)\frac{\partial{n}}{\partial{x}}]I
                                    \tag{31}
\end{align*}
Comparing Eq.(31) with Eq.(30) and noting that the coefficient of $\frac{\partial^2{n}}{\partial{x^2}}$ should be the same, $g(x)$ is obtained as:
\begin{align*}
g(x)=-\frac{N_e}{2}(\frac{1}{KT_e})^2I=-Ax^2(1+Bx)\nn\tag{32}
\end{align*}
Where $A=\frac{KT_e}{m_ec^2}N_e\sigma_Tc$, $B=\frac{14}{5}\frac{KT_e}{m_ec^2}$. Inserting Eq.(32) into the Eq.(29), we obtain Eq.(33),
\begin{align*}
(\frac{\partial{n}}{\partial{t}})_c=&\frac{KT_e}{m_ec^2}N_e\sigma_Tc\frac{1}{x^2}\frac{\partial}{\partial{x}}\{x^4(1+\frac{14}{5}\frac{KT_e}{m_ec^2}x)[\frac{\partial{n}}{\partial{x}}+n(n+1)]\}\tag{33}
\end{align*}

Eq.(33) is a new equation. Here we should point out that the we use $\Delta p$ rather than $\Delta \nu$ in the Taylor expansion.
This is the main reason that our result is different from the previous results.
The new equation can be applied in the nonrelativistic energy regime with the photon energy $h\nu << m_ec^2$ and
the electron temperature $KT_e << m_ec^2$ to describe a more general Compton Scattering process.
The comparison between $h\nu$ and $KT_e$ is no longer necessary. When the energy of photons $h\nu$ is low,
then the term $\frac{14}{5}\frac{KT_e}{m_ec^2}x$ of Eq.(33) can be ignored , so it return to the
classical Kompaneets equations. While when the energy of photons $h\nu$ is very high,
the effect of the term $\frac{14}{5}\frac{KT_e}{m_ec^2}x$ of Eq.(33) could have large effect
and hence is not negligible.

\section{Numerical Calculations}

We apply our new equation to explore evolution of typical X-ray spectra in astro-physics.
We calculate Eq.33 numerically by using the three-point finite difference method.
We fix $N_e=2\times10^{16}cm^{-3}$, which are the typical values for plasma in the accretion
disk around a compact star.

The expression of the spectral intensity for an emission line
with normal Gaussian profile is given as:
\begin{align*}
     I(v)\sim exp[-\frac{4ln2}{(\Delta \nu)^2}(\nu - \nu_0)^2]\tag{34}
\end{align*}

Using the relation $I_v = \frac{2h\nu_3}{c^2}n(\nu)$, The initial line profile of a stable source is given as:
\begin{align*}
     n(x,0)=f(x)\sim x^{-3}exp[-\frac{4ln2}{(\Delta x)^2}(x-x_0)^2]\tag{35}
\end{align*}

\begin{figure}[htbp]
\centering
\includegraphics[width=0.4\textwidth]{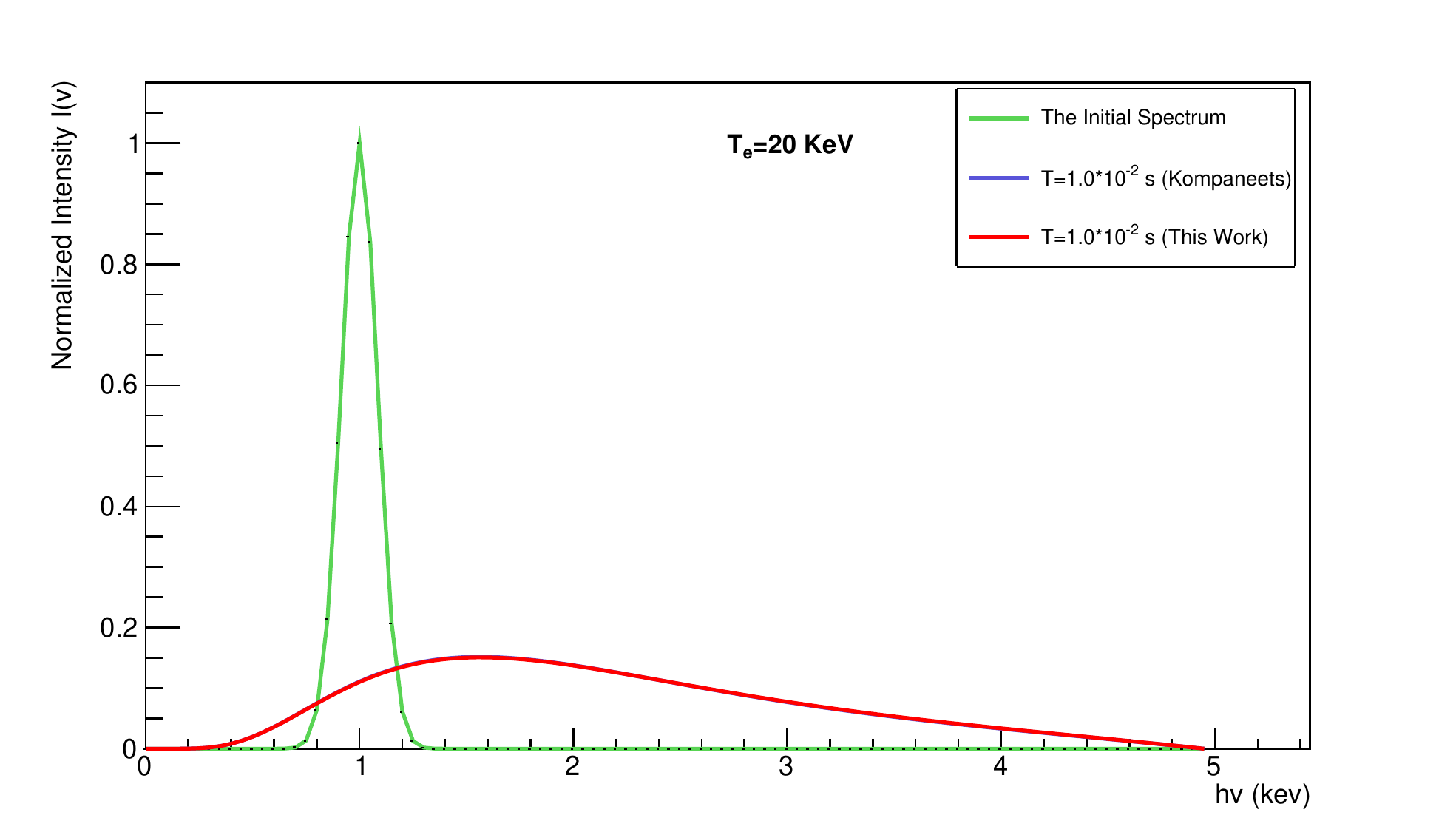}
\includegraphics[width=0.4\textwidth]{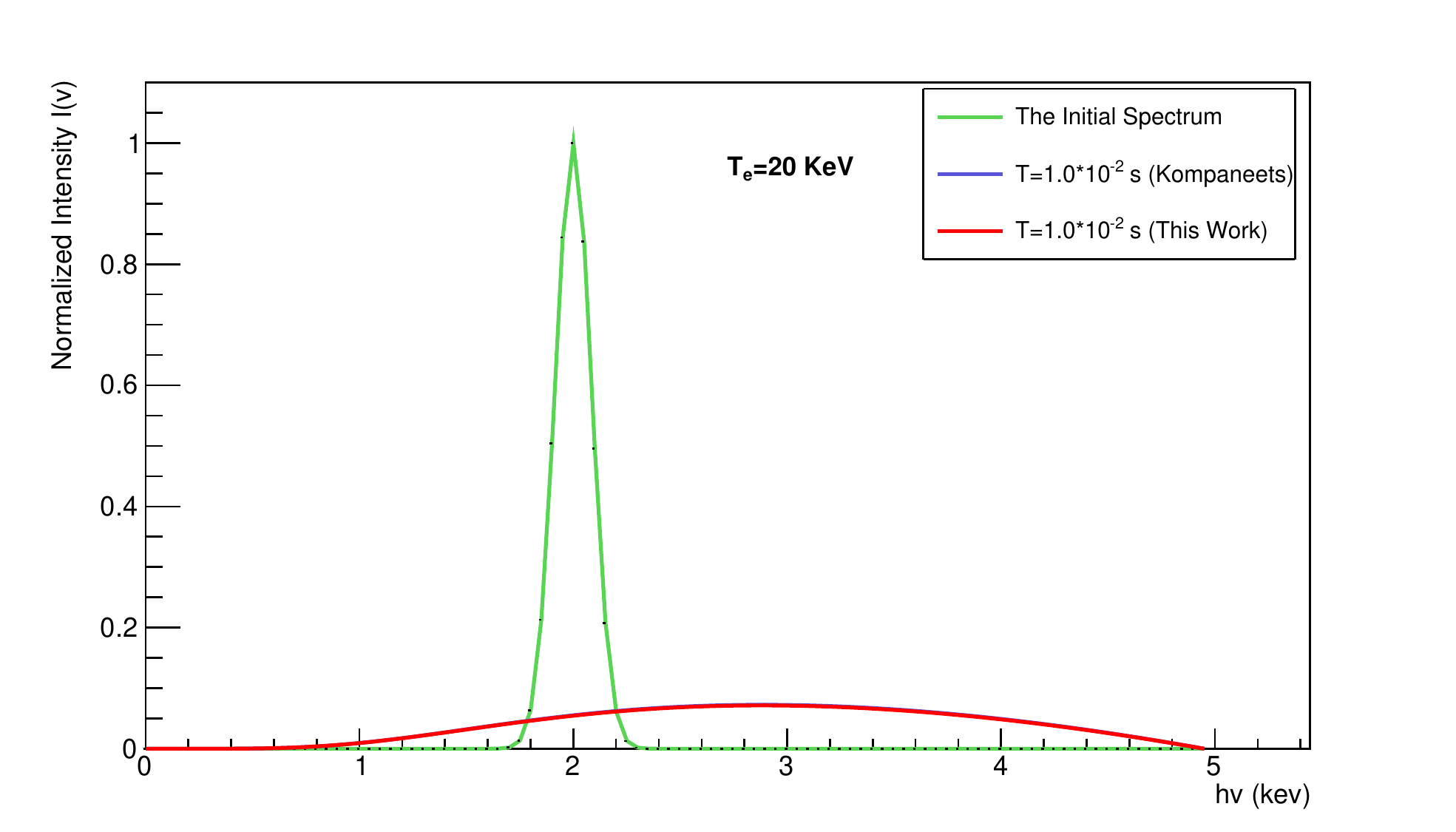}
\caption{(Color Online) Left:$h\nu_0 = 1.0$ KeV, the FWHM of the line as $h\Delta\nu$  = 0.1 KeV, 0.001 KeV $< h\nu <$ 5 KeV.
Right: $h\nu_0$ = 2.0 KeV,the FWHM of the line as $h\Delta \nu$ = 0.2 KeV, 0.001 KeV $< h\nu <$ 5 KeV.}
\label{fig:fig12}
\end{figure}

In the following calculations, we take the diffusion time scales is $T = 1.0\times 10^{-2}$.
Fig~\ref{fig:fig12} shows our new equation compared with the Kompaneets equation in up-Comptonization process of low energy photons passing
through electron plasma. We can see the difference of the resultant line-profiles between our work and Kompaneets
is very small, which confirms that the correctness of Eq. (33) is valid in up-Comptonization process.

\begin{figure}[htbp]
\centering
\includegraphics[width=0.4\textwidth]{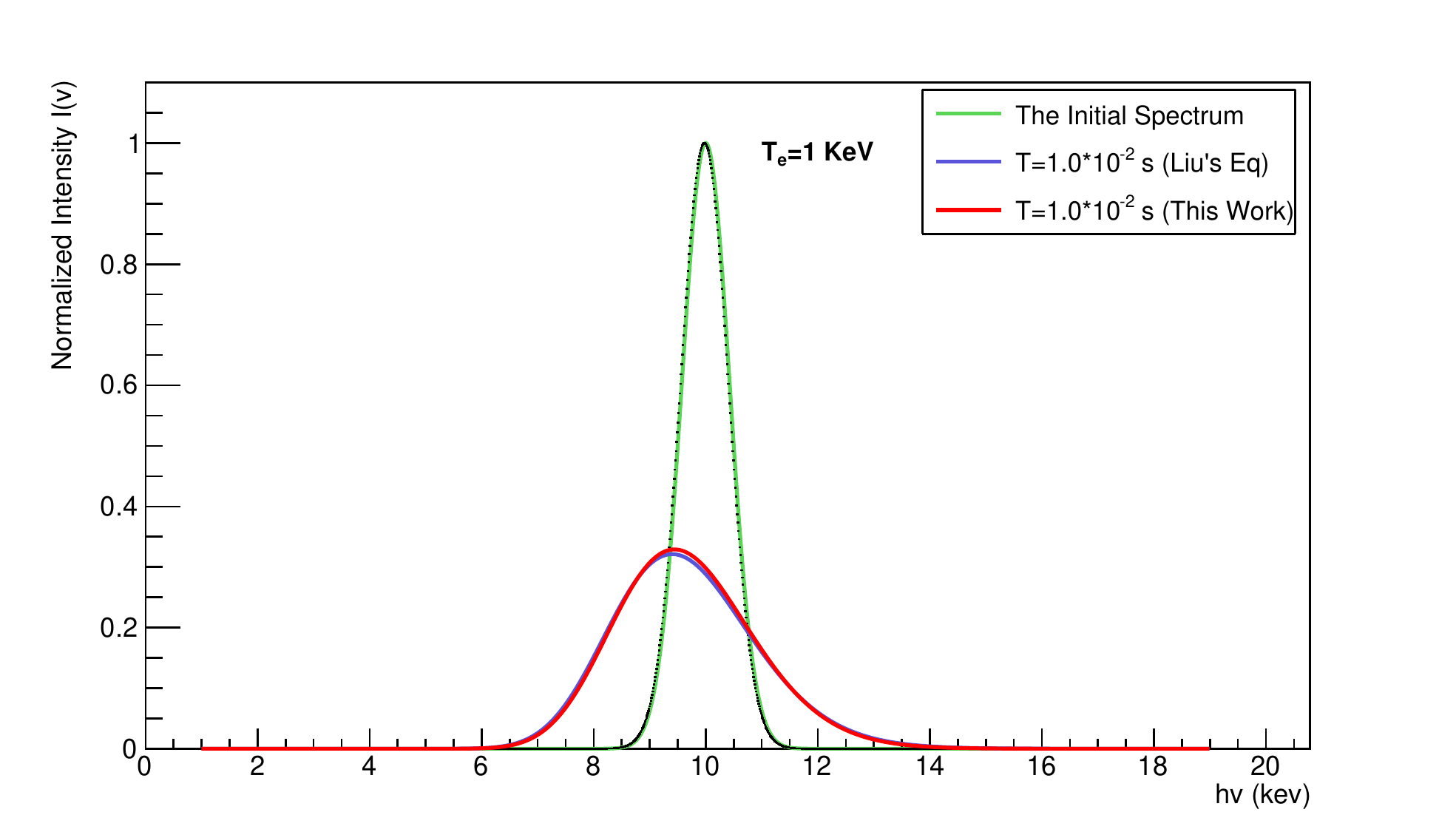}
\includegraphics[width=0.4\textwidth]{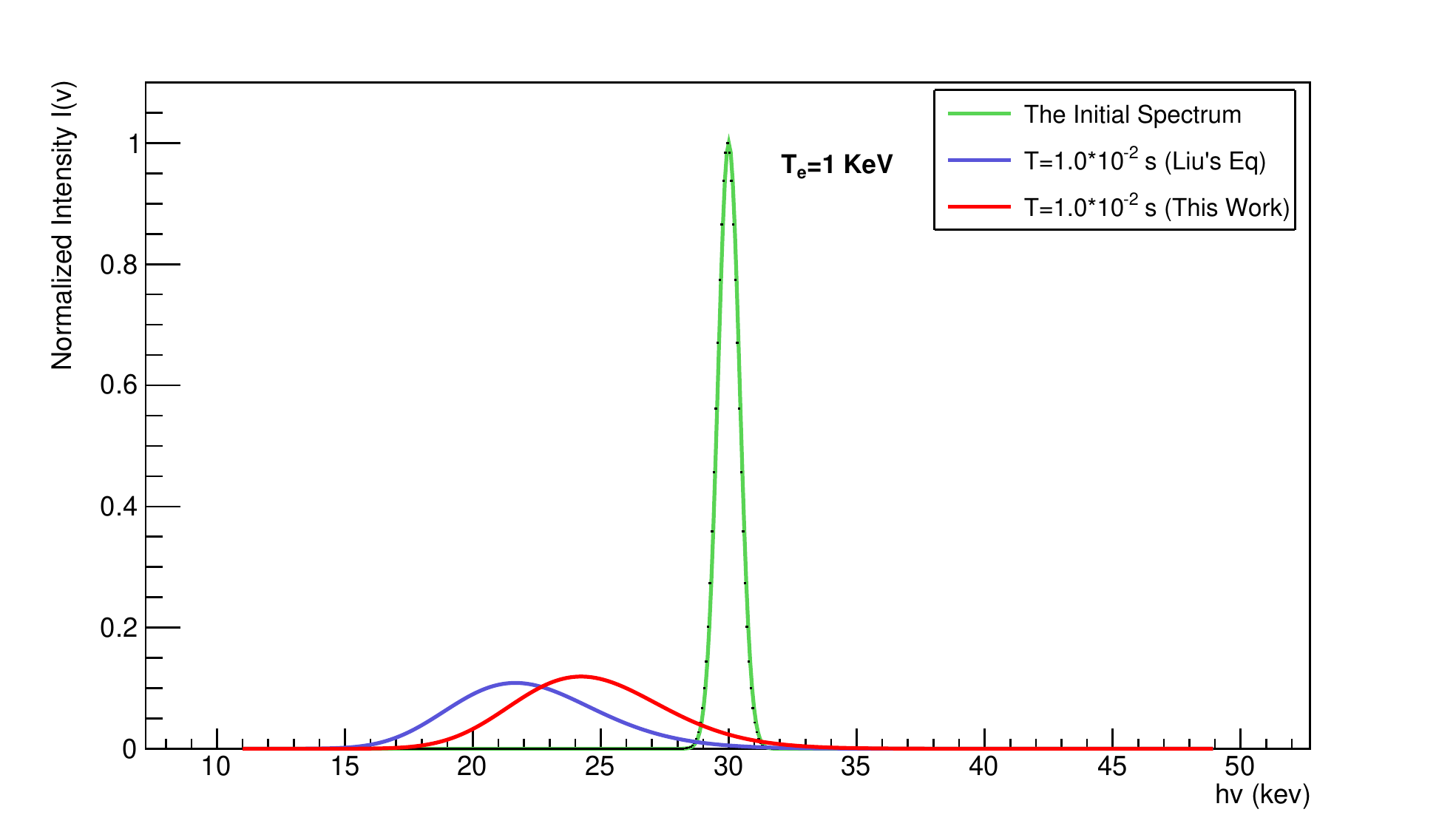}
\includegraphics[width=0.4\textwidth]{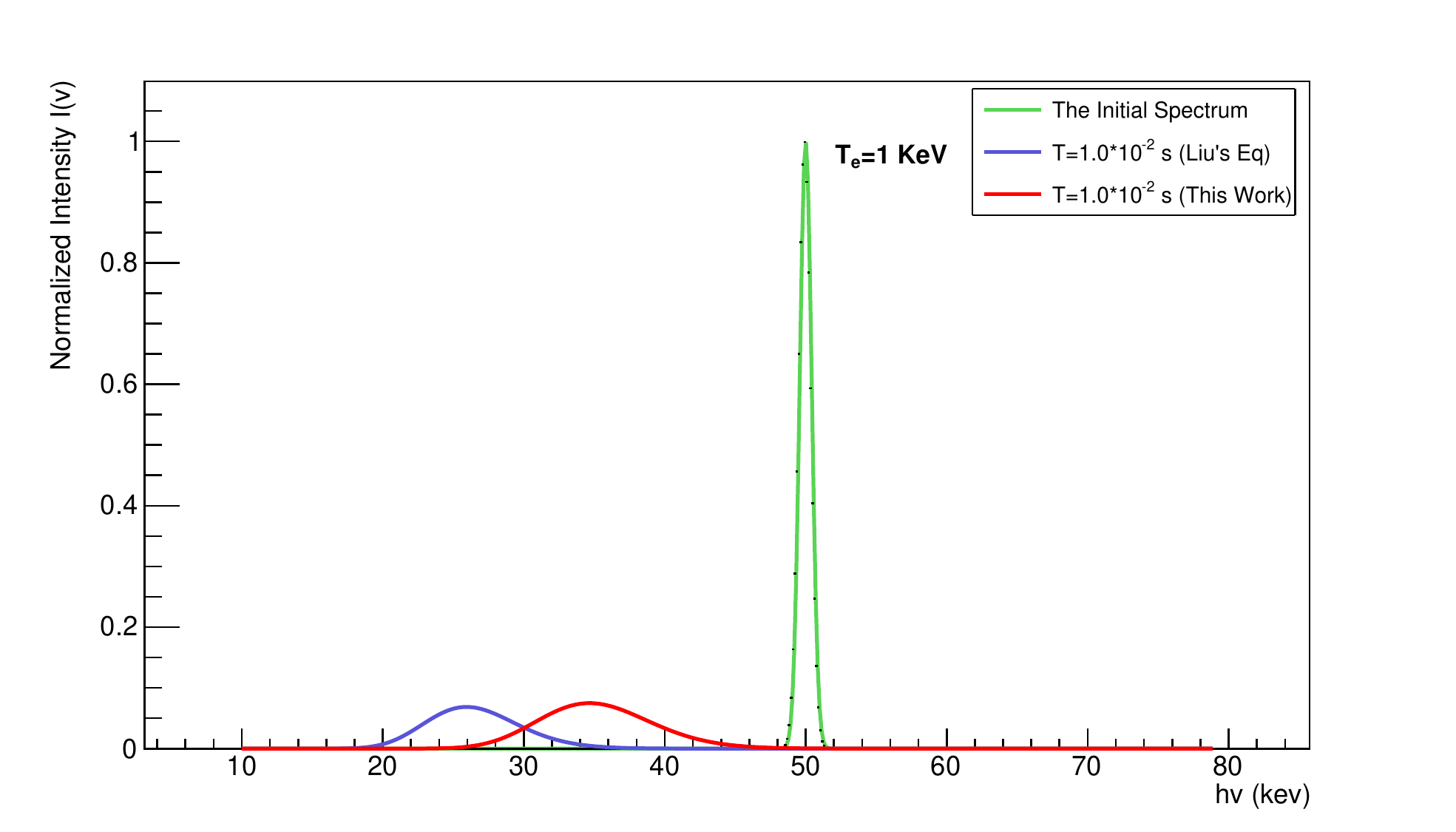}
\caption{(Color Online)
Top Left: $h\nu_0$ = 10.0 KeV,the FWHM of the line as $h\Delta\nu$ = 1.0 KeV, 1.0 KeV $< h\nu <$ 19KeV.;
Top Right: $h\nu_0$ = 30.0 KeV,the FWHM of the line as $h\Delta\nu$ = 1.0 KeV, 11.0 KeV $< h\nu <$ 49.0KeV;
Bottom: $h\nu_0$ = 50.0 KeV,the FWHM of the line as $h\Delta\nu$ = 1.0 KeV, 11.0 KeV $< h\nu <$ 79.0 KeV.}
\label{fig:fig345}
\end{figure}

For down-Comptonization process, the Kompaneets equation does not work any longer.
Here we compare our equation with Liu's equation. Fig.~\ref{fig:fig345} shows our work compared
with Liu's work in down-Comptonization process of high energy photons passing through electron plasma.
From Fig.~\ref{fig:fig345}, we conclude that when the centroid energy as $h\nu_0$ is not very large,
the difference between our equation and Liu's equation is very small.
while with the increasing of $h\nu_0$ , the difference is more and more obvious.
As we know if higher order term of the distribution function is ignored,
the small quantity is smaller, the result is more precise. When $h\nu_0$ is not very large, the second
and higher order term of the distribution function expanded by $\Delta p$ is approximately to that of
expanded by $h\Delta \nu$. While when $h\nu_0$ is very large, the obvious difference between our equation
and Liu's equation comes from the second and higher order term of the distribution function which is
expanded by $\Delta p$ is much smaller than that expanded by $\Delta\nu$.

\section{Conclusions}

In this study, we have explored both up-Comptonization and down-Comptonization processes
using relativistic corrections to Kompaneets equation to describe a more general Compton
Scattering process. While different from Kompaneets derived his equation, we have choose $\Delta{p}$
as the small quantity. When $h\nu >> KT_e$, $\Delta{p}$ is more approximately to the small quantity
than $\Delta{\nu}$, which ensure the equations is more accurete for down-Comptonization processes.
In up-Comptonization process, our results match the Kompaneets equation. With introducing
relativistic corrections to Kompaneets equation, the non-relativistic process
with photon energy $h\nu << m_ec^2$ and electron temperature $KT_e << m_ec^2$
can be described, with the conditions $h\nu << KT_e$ is not necessary.

{\bf Acknowledgments}
We thank Dangbo Liu and Jiajie Ling for fruitful discussion and suggestion.
We are grateful Chengdong Han help us made the plots.
This work is partly supported by the national 973 program 2014CB845406, the National Natural Science Foundations of
China under the Grants Number 11175220 and Century Program of
Chinese Academy of Sciences Y101020BR0.

\clearpage

\end{document}